\documentclass[conference]{IEEEtran}
\IEEEoverridecommandlockouts
\usepackage{cite}
\usepackage{amsmath,amssymb,amsfonts}
\usepackage{algorithmic}
\usepackage{graphicx}
\usepackage{textcomp}
\usepackage{xcolor}
\usepackage{comment}
\usepackage{algorithm}
\usepackage{graphicx}
\usepackage{textcomp}
\usepackage{xcolor}
\usepackage{subcaption}
\usepackage{xspace}
\usepackage{hyperref}
\usepackage{listings}
\usepackage{booktabs}
\usepackage{tabularx}
\usepackage{multirow}
\usepackage{booktabs}
\usepackage{colortbl}
\usepackage{tikz}
\usepackage{ctable}
\usepackage{listings}

\usepackage[skip=2pt,font=small]{caption}
\usepackage{balance} 
\usepackage{nicematrix}
\usepackage{textcomp}
\usepackage{pifont}
\usepackage{float}
\usepackage{stfloats}
\newcommand{\squishlist}{
   \begin{list}{$\bullet$}{%
        \setlength{\itemsep}{0pt}%
        \setlength{\parsep}{0pt}%
        \setlength{\topsep}{0pt}%
        \setlength{\partopsep}{0pt}%
        \setlength{\listparindent}{-2pt}%
        \setlength{\itemindent}{-5pt}%
        \setlength{\leftmargin}{1.2em}%
        \setlength{\labelwidth}{0em}%
        \setlength{\labelsep}{0.5em}%
    }
}

\newcommand{\squishend}{
    \end{list}  }

\newcommand{\circlednumber}[1]{%
    \begin{tikzpicture}[baseline=(C.base)]
        \node[minimum size=0.35cm,shape=circle,draw,inner sep=0.5pt,fill=black,text=white,font=\bfseries] (C) {#1};
    \end{tikzpicture}%
}

\newcommand {\nickname}{\textsc{MAGE}\xspace}

\def\BibTeX{{\rm B\kern-.05em{\sc i\kern-.025em b}\kern-.08em
    T\kern-.1667em\lower.7ex\hbox{E}\kern-.125emX}}
\begin{document}

\title{MAGE: A Multi-Agent Engine for Automated RTL Code Generation}

\author{\IEEEauthorblockN{
Yujie Zhao\textsuperscript{*},
Hejia Zhang\textsuperscript{*},
Hanxian Huang,
Zhongming Yu,
Jishen Zhao}
\IEEEauthorblockA{
University of California San Diego\\
La Jolla, CA 92093, USA\\
\{yuz285, hez024, hah008, zhy025, jzhao\}@ucsd.edu}
\thanks{* Equal Contribution}
}

\maketitle

\begin{abstract}
The automatic generation of RTL code (e.g., Verilog) through natural language instructions has emerged as a promising direction with the advancement of large language models (LLMs). 
However, 
producing RTL code that is 
both syntactically and functionally correct remains a significant challenge. 
Existing single-LLM-agent 
approaches face substantial limitations because they must navigate between various programming languages and handle intricate generation, verification, and modification tasks.
To address these challenges, this paper introduces \nickname, the first open-source\footnote{\nickname is open-sourced at \url{https://github.com/stable-lab/MAGE-A-Multi-Agent-Engine-for-Automated-RTL-Code-Generation}.} multi-agent AI system designed for robust and accurate Verilog RTL code generation. 
We propose a novel high-temperature RTL candidate sampling and debugging system that effectively explores the space of 
code candidates and significantly improves the quality of the candidates. 
Furthermore, we design a novel Verilog-state checkpoint checking mechanism that enables early detection of functional errors and
delivers precise feedback for targeted fixes, significantly enhancing the functional correctness of the generated RTL code. 
\nickname achieves a 95.7\% rate of syntactic and functional correctness code generation on VerilogEval-Human v2 benchmark, surpassing the state-of-the-art Claude-3.5-sonnet by 23.3\%, demonstrating a robust and reliable approach for AI-driven RTL design workflows. 
\end{abstract}


\section{introduction}\label{sec:intro}

Digital hardware design usually requires engineers to define the architecture and functionality of hardware by writing code in hardware description languages (HDLs), such as Verilog and VHDL. 
%
As VLSI designs become more complex, hardware design workflows using HDLs are increasingly time-consuming and error-prone~\cite{dessouky2019hardfails,ahmad2024hardware,rtl-repair}, often requiring multiple iterations to debug functional correctness.
%
Although electronic design automation (EDA) tools have advanced to support this workflow~\cite{lavagno2018eda,scheffer2018eda,wakabayashi2000c,nane2015survey}, the demand for more efficient and automated hardware design solutions remains crucial.

%
%

Large Language Models (LLMs) ~\cite{achiam2023gpt, Claude3} have recently shown
promising potential in assisting and improving hardware design~\cite{thakur2023benchmarking, pei2024betterv, tsai2024rtlfixer,cui2024origen}.
%
%
%
Recent studies explored leveraging LLMs to improve the correctness of RTL code generation through model fine-tuning~\cite{pearce2020dave,thakur2023benchmarking,liu2024rtlcoder}, training of domain-adaptive models~\cite{liu2023chipnemo}, and single-agent methods~\cite{huang2024towards, gao2024autovcoder, cui2024origen} that augment LLMs by incorporating results and feedback from compilers or simulators to refine the generated code. 
%
%
%
%

%
However, previous studies using a single LLM or a single agent face substantial limitations. First, these approaches fall short of the context-switch between tasks, programming languages, and knowledge domains. 
%
As a result, the quality of RTL design suffers: for instance, a single Claude-3.5-sonnet agent~\cite{Claude3} has a functionality pass rate of only $75.0\%$ even for simple design tasks~\cite{lu2024rtllm,liu2023verilogeval,pinckney2024revisiting}. 
Second, HDLs are specifically designed to describe the logic and architecture of digital hardware at a low level, focusing on the flow of data between registers, the timing of operations, and hardware characteristics such as propagation delays and signal dependencies. Directly adopting general-purpose LLM or multi-agent designs will produce low-accuracy RTL code that fails to meet timing constraints, resulting in wrong designs~\cite{lu2024rtllm,liu2023verilogeval,thakur2023benchmarking,liu2023chipnemo}. 
%
%
%

To address these challenges, we propose \textbf{\nickname}, the first open-source \textbf{M}ulti-\textbf{AG}ent \textbf{E}ngine to achieve automatic high-quality RTL code generation. 
Different from the existing single-agent RTL code generation frameworks~\cite{huang2024towards, gao2024autovcoder, cui2024origen}, \nickname enables specialized agents to handle distinct aspects in the RTL development pipeline. 
Inspired by real-world RTL design workflows, where specialists focus on distinct stages, our approach, \nickname, employs a specialized multi-agent architecture composed of four types of key agents: the RTL code generation agent, testbench generation agent, judge agent, and debug agent. Each type plays a specific role, working collaboratively within our tailored recursive framework to generate optimized and reliable RTL code. 

\nickname consists of three key \textbf{design principles}. First, mimicking the iterative nature of human RTL design teams in addressing complex design challenges, we developed a high-efficiency collaboration workflow with a delicate design context communication protocol. Second, we propose a novel high-temperature RTL candidate sampling and debugging system, which leverages both high-temperature sampling and simulation-based scoring to identify promising candidates. Our key insight is that higher temperatures result in more diverse LLM outputs, while also increasing the likelihood of including the highest-quality candidates for optimization in the next stage. Finally, we propose a novel Verilog-state checkpointing and validation scheme, which substantially improves the functional correctness. Unlike conventional methods that provide feedback only on the final output mismatch~\cite{huang2024towards, cui2024origen,gao2024autovcoder}, \nickname validates an RTL design by comparing state values with expected outputs at each clock edge. This enables early detection of functional errors and provides precise feedback for targeted fixes.

In summary, we make the following contributions: 
\squishlist
\item 
We design \nickname, the first open-source LLM-based
multi-agent system for robust and accurate Verilog RTL code generation. By decomposing a complex hardware design into manageable sub-tasks, we design LLM agents to handle sub-tasks and design the system to enable effective agent communication and collaboration. 
\item 
We propose a novel high-temperature RTL candidate sampling system, notably enhancing generated code quality
by exploring the benefits of high-randomness
generation.
\item 
We propose a novel
Verilog-state checkpoint checking mechanism, which provides precise feedback for targeted fixes, significantly improving the RTL code quality.
\item 
\nickname achieves a \textbf{95.7\%} rate of syntactic and functional correctness code generation on the VerilogEval-
Human v2 benchmark, surpassing all existing methods, representing a critical step toward automating and optimizing hardware design workflows, offering a more robust methodology for AI-driven RTL design.
\squishend

\section{Background and Motivation}
\subsection{Characteristics and challenges of LLM-based RTL design}
\label{background}
\noindent\textbf{Decomposing Complex Traditional RTL Design.}
Traditional digital hardware design flows necessitate hardware engineers to iteratively perform: 
(1) implement Verilog code 
to specify hardware architectures and behaviors, (2) 
customize test benches to rigorously verify the correctness of these hardware descriptions (3) iterate between Verilog simulations, signal waveform reasoning, and code refining until all output signals match expected behavior. 
This iterative loop among design, verification, and refinement makes the RTL design process not only challenging but also highly demanding in terms of both time and expertise. The complexity of traditional Verilog RTL design calls for decomposing the whole process into multiple manageable sub-stages and adopting different specialists (agents) for different sub-stages.   

\noindent\textbf{Related Work on LLM-based RTL Design.}
Recent works \cite{liu2023chipnemo,thakur2023benchmarking,liu2024rtlcoder, wu2024itertl,zhao2024codev} train or fine-tune general code generation LLMs by incorporating RTL domain knowledge to enhance the correctness of RTL code generation. 
Single-agent methods~\cite{huang2024towards,cui2024origen,gao2024autovcoder} further integrate simulation results and introduce more stages in the code generation process, such as planning, verification, and code refining based on the feedback from simulation. However, this process involves generating both synthesizable and non-synthesizable code for RTL generation and verification, and context switches among different sub-tasks in multiple iterations, necessitating different domain knowledge and problem-solving abilities. Thus, adopting a single agent for all these tasks in the hardware design process leads to sub-optimal results, as they have to process such complex contexts in each interaction and maintain consistency through a long unified conversation history.

In contrast, multi-agent systems distribute tasks among agents with independent conversation histories, enabling specialized task handling and greater modularity. However, 
Aivril~\cite{sami2024aivril} only implements a basic two-agent division between code generation and review, still requiring a single agent to handle both synthesizable RTL and non-synthesizable verification code, thus failing to address the fundamental context-switching challenges.
Verilogcoder~\cite{ho2024verilogcoder} further constrains accessibility and system adaptability through its closed-source implementation and reliance on proprietary components (e.g., its Abstract Syntax Tree and Waveform Tracing Tool).

\noindent\textbf{Temperature Sampling for LLM-based RTL Design.}  Given an input requirement, LLMs rely on a specific decoding strategy to generate the code auto-regressively. Temperature sampling method~\cite{ackley1985learning} uses a temperature coefficient $T$ (usually $\in[0,1]$) to control the sampling randomness. 
Increasing the temperature avoids overly conservative results and promotes diversity in code generation, thus enhancing the chance of exploring the correct answers. However, this comes at the cost of introducing more noise and errors in the generation results. Recent studies \cite{zheng2023codegeex, zhu2024hot} show that high-temperature sampling can improve correctness in software code generation through multiple sampling iterations. However, higher temperatures tend to result in poorer performance for RTL design as explored in recent studies \cite{thakur2023benchmarking, pinckney2024revisiting}. We found a key reason for this is that single-agent mechanisms restrict the ability to design independent and efficient sampling and feedback, thereby hindering the optimization potential of high-temperature sampling for RTL code.

\begin{figure*}[ht]
    \centering
    \includegraphics[width=1\linewidth]{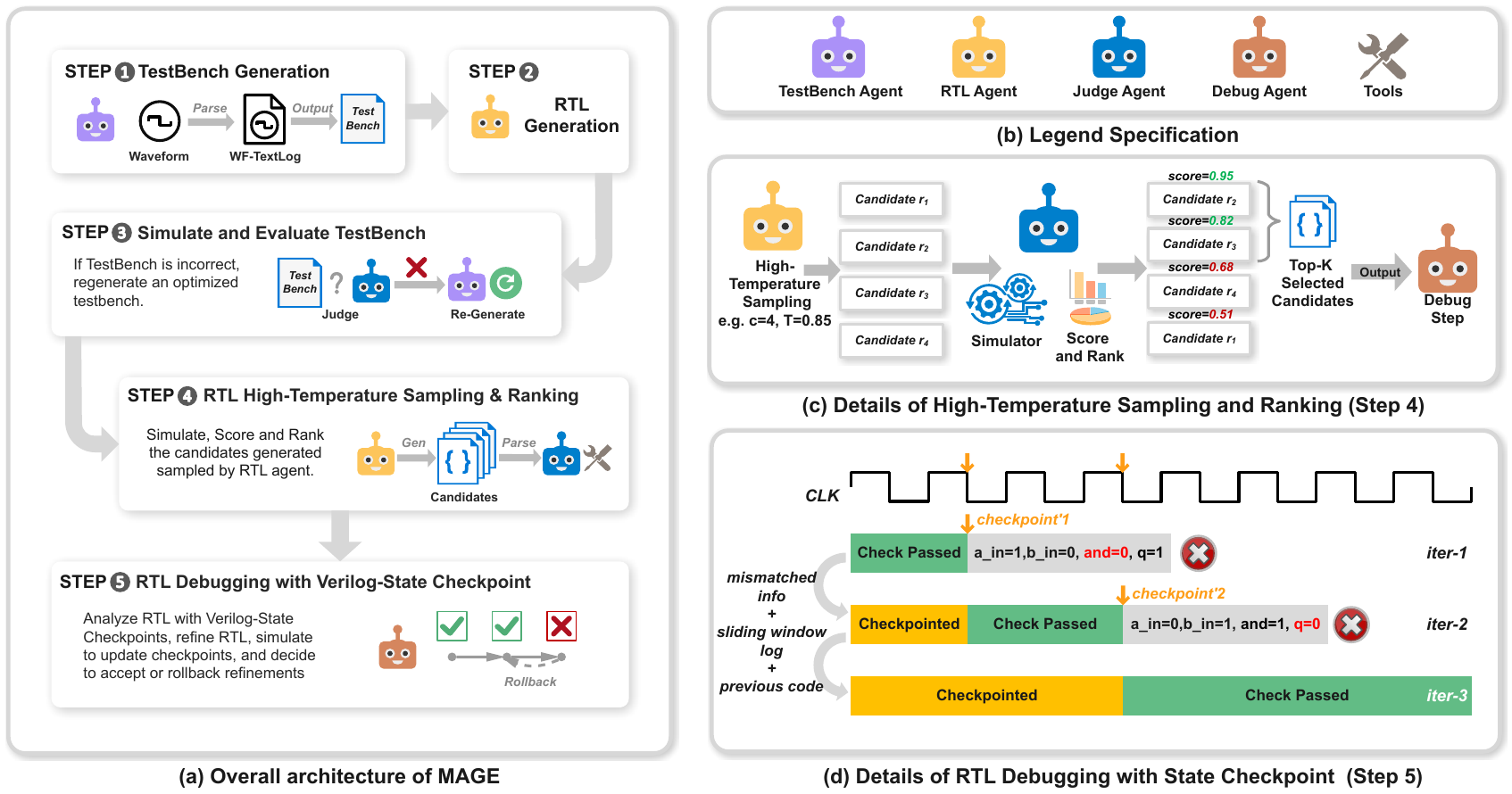}
   
    \caption{(a) The overview of {\nickname}, (b) the roles of four types of agents, (c) state check module, and (d) sampling and debugging module. }
    \label{fig:fig-workflow}
    \vspace{-15pt}
\end{figure*}

\subsection{Opportunity on Effective Code Generation}
\label{opportunity-sampling}
To tackle the aforementioned limitations of both existing single-agent and multi-agent systems for RTL code generation, we identify potential opportunities to develop effective RTL code generation frameworks to improve correctness. 

First, to further reduce the mutual influence between non-synthesizable testbenches and synthesizable RTL code generation, as well as to decrease the complexity of tasks assigned to each agent, we recognize the necessity of appropriately distributing tasks among multiple agents and organizing the workflow effectively. Specifically, previous works (e.g., \cite{huang2024towards, sami2024aivril}) employed a single agent to generate both non-synthesizable testbenches and synthesizable RTL code simultaneously. However, \cite{huang2023agentcoder} found that generating testbenches and code together within the same conversation may lead to a corresponding decrease in effectiveness. Moreover, test cases generated in this manner can become biased and influenced by the code, resulting in a loss of objectivity and diversity.

Second, based on our analysis in Sec.~\ref{background}, we thoroughly reviewed existing multi-agent systems\cite{sami2024aivril,ho2024verilogcoder}, where we found that all of them are closed-source and dependent on third-party proprietary tools that are not LLM-directly-adapted. This inevitably limits system extensibility and transparency. Therefore, we identify the need for an open-sourced multi-agent system.

Finally, to address the limitation of high-temperature sampling for RTL code generation described in Sec.~\ref{background}, we summarize the key to efficiently sampling RTL code candidates process to mitigate the impact of randomness while leveraging the exploration benefits. Given that RTL code is synthesizable and a practical miscount scoring method is applicable, we prioritize candidate selection at an early stage by identifying top-scoring candidates. This approach allows us to reduce the exploration cost by filtering suboptimal results early.


\subsection{Opportunity on Effective Code Debugging}
\label{opportunity-debugging}
Most existing LLM approaches for RTL code generation~\cite{cui2024origen,huang2024towards,gao2024autovcoder} rely on the direct application of original golden testbenches, which can only provide pass rates. This provides very limited feedback to LLMs in fine-tuning RTL code as pass rates alone lack detailed insights into crucial aspects such as timing analysis, signal interactions, variable values, and mismatch information, all of which are essential for ensuring high-quality RTL design. VerilogCoder~\cite{ho2024verilogcoder} applies closed-source Abstract Syntax Tree (AST) analysis, which significantly restricts flexibility and transparency. Furthermore, the introduction of tools that output in graphical form, which LLM can not directly apply, increases the complexity of tasks for LLMs and reduces their effectiveness.  Therefore, we identify the requirement for an optimized testbench, which will output a log resembling a simulated waveform in text form, which can be directly adaptable by LLMs. This shifts the reliance on closed-source tools to a fully open-source LLM-based textual protocol, alongside significant improvements in LLM RTL code debug effectiveness, improving both the extension of the framework and the quality of the analysis.

\section{\nickname Design}

We present \nickname, a multi-agent engine designed specifically for RTL. This system integrates an RTL-specific context communication protocol with high-efficiency, LLM-adapted tools for fine-tuning RTL, all within a productive and orchestrated process.

\subsection{Multi-Agent System Overview}

Figure~\ref{fig:fig-workflow} depicts a comprehensive overview of the workflow implemented in \nickname. Inspired by the collaborative division of work approaches of human RTL design teams, our framework incorporates four types of agents, each playing a specific and concrete role in the automation process. Figure~\ref{fig:fig-workflow} (b) specifically delineates the responsibilities and tasks assigned to each agent, ensuring a clear understanding of the workflow. (1) A \underline{Testbench Generation Agent} is responsible for creating optimized test benches in a textual-waveform-output format based on the natural language specifications and any available golden test benches. (2) A \underline{RTL Generation Agents} convert these specifications along with the optimized test bench into Verilog code, incorporating syntax checking to ensure code validity. (3) A \underline{Judge Agent} then takes over by simulating and evaluating the generated RTL code against the optimized test bench. It scores the RTL code candidates and decides whether any require debugging or if the test bench needs to be regenerated. (4) A \underline{Debug Agent} performs iterative refinements on any code that fails initial tests, using textual waveform-like simulation outputs as feedback for improvements. Our method mimics the division of labor in human teams and enhances the efficiency and accuracy of automated RTL design. It should be noted that, whether in RTL code generation or debugging, the agent will perform at most $s=5$ iterations to automatically fix syntax errors.

Based on the distinct and specific roles of multiple agents, we have designed an efficient workflow to ensure that each agent's unique contributions are seamlessly integrated, facilitating a coherent progression throughout the workflow. Figure~\ref{fig:fig-workflow} (a) demonstrates the entire process is systematically divided into five main steps as depicted as follows:

\noindent Step~\circlednumber{1} -- Generate initial textual-waveform-output testbenches.  
To reduce the limitations of pass-rate-output golden testbenches, which is analyzed in Sec.~\ref{opportunity-debugging}, we directly utilize natural language specifications to generate optimized testbenches that can output State Checkpoints, which will be used in Step \circlednumber{5}  (see Figure~\ref{fig:fig-workflow} (d) and Sec.~\ref{debugging-details}). Furthermore, since natural language specifications may contain ambiguities, we combine the input with the golden testbench if available. 

\noindent Step~\circlednumber{2} -- Based on the natural language specifications and the optimized testbench, we generate an initiate RTL code. 

\noindent Step~\circlednumber{3} -- If the initial RTL code cannot pass the optimized testbench, the judge agent evaluates the testbench and regenerates it if deemed incorrect. 

\noindent Step~\circlednumber{4} -- If the RTL code is deemed correct, we will employ an \textbf{High-Temperature RTL Sampling and Scoring Process} (see Figure~\ref{fig:fig-workflow} (c) and detailed in Sec.~\ref{samling-details}) to generate and ranking RTL code Candidates. 

\noindent Step~\circlednumber{5} -- If the RTL code candidates still have function errors, we will employ a \textbf{RTL Debugging with State Checkpoint Mechanism} (see Figure~\ref{fig:fig-workflow} (d) and Sec.~\ref{debugging-details}) to make effective debug trials and debug the selected RTL code candidates.

\subsection{High-Temperature RTL Sampling and Scoring}

\begin{figure}[ht]
    \centering
    \includegraphics[width=\linewidth]{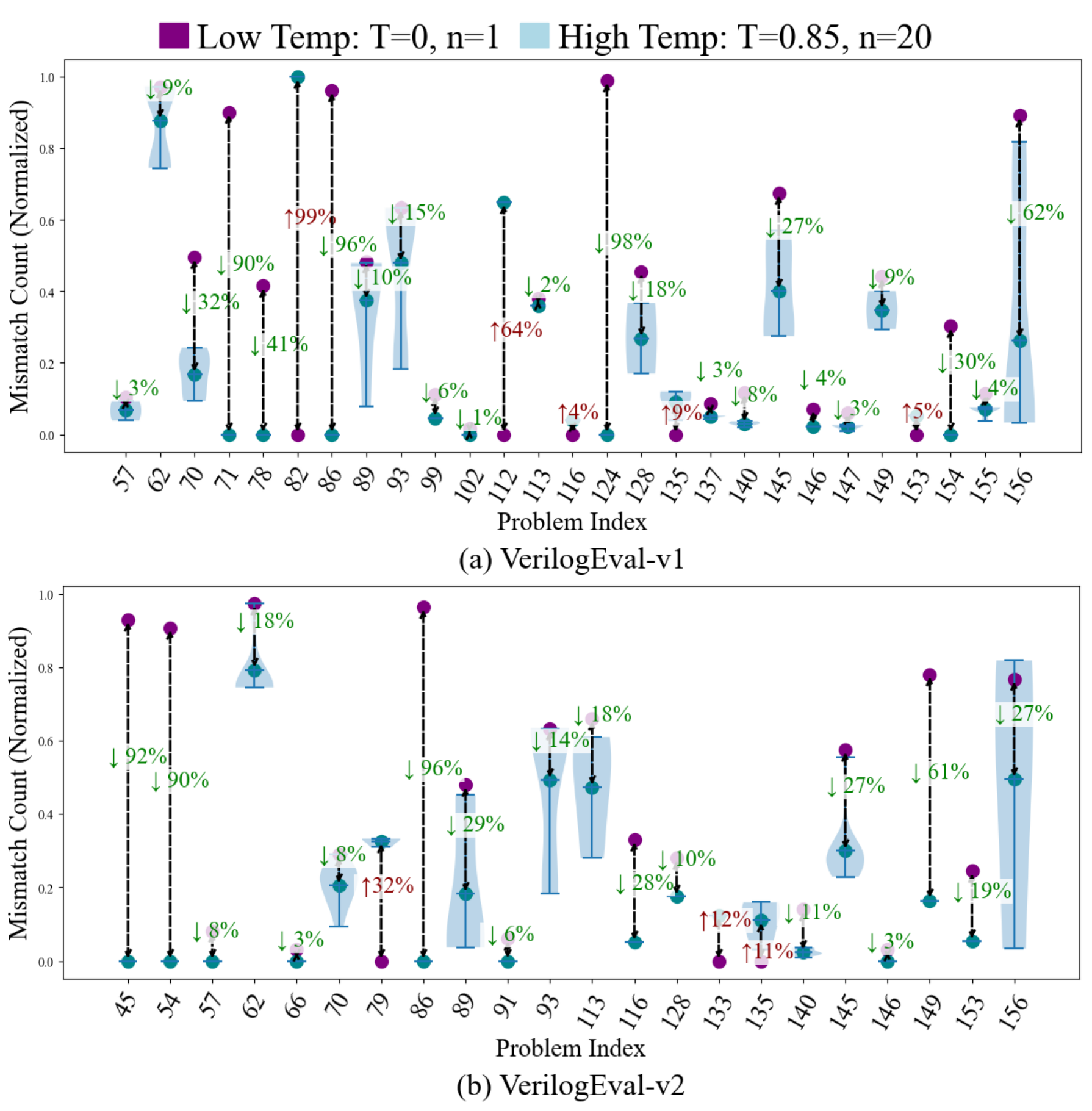}
   
    \caption{Normalized mismatch count of generated testbenches at different stages under varying temperature configurations (Low temperature: $T=0$, $n=1$; High temperature: $T=0.85$, $n=20$), using the Claude 3.5 Sonnet model (dated 2024-10-22) across two benchmarks: VerilogEval-v1-Human~\cite{liu2023verilogeval} and VerilogEval-v2~\cite{pinckney2024revisiting}. Problems that directly passed before Step \protect\circlednumber{4} and those with zero mean mismatches in both configurations are not shown in the figure. The blue violin plot shows that candidates (blue dots) generated with high-temperature sampling typically have lower mean mismatch counts across most problems compared to those generated with low temperatures (purple dots).} 
    \label{fig:fig-sampling-ablation}

\end{figure}
\label{samling-details}
As depicted in Sec.~\ref{opportunity-sampling}, high-temperature sampling has a better chance of exploring the correct code. However, in prior studies on Temperature Sampling for RTL code generation like \cite{zheng2023codegeex, zhu2024hot}, it is commonly acknowledged that high-temperature sampling introduces significant randomness, often resulting in reduced correctness. To address this issue, we first explore the potential opportunity of using multiple RTL code generation candidates in high-temperature sampling to reduce simulated mismatch counts in one iteration. 
  Fig. \ref{fig:fig-sampling-ablation} shows the distribution of normalized mismatch counts for the best candidates at two different temperatures. We observe that the best candidate with high temperature has a lower mismatch count for most problems. This suggests that, despite increased randomness, high-temperature sampling yields better performance when the sampling is sufficiently extensive. The improvement in pass rates at each stage, as shown in Fig. \ref{fig:fig-sampling-ablation}, indicates the potential benefits of performing an effective sampling policy at high temperatures. 


Based on the above observation, we propose a High-Temperature Sampling and Ranking Process as follows:

\noindent1. We sample $c$ RTL code candidates for issue $i$: 
\begin{align}
 \mathcal{R}_i = \{r_{i}^{1}, r_{i}^{2}, \ldots, r_{i}^{c}\} , r_{i} \sim P_{T}\left(r \mid p_{sys}, SP_i, TB_i\right)
\end{align}
Here, $T$ is the temperature, $p_{sys}$ denotes the system prompt of the RTL code generation, $SP_i$ is the natural language specification of the issue $i$, and $TB_i$ is the corresponding testbench. The term $P_{T}\left(r \mid p_{sys}, SP_i, TB_i\right)$ represents the probability distribution of the generated RTL code given the temperature, system prompt, specification, and testbench.

\noindent2. We select the Top-$K$ candidates from $\mathcal{R}_i$ based on their scores. Specifically, each candidate in $\mathcal{R}_i$ is evaluated to obtain a normalized mismatch-based score as follows:
\begin{align}
    s(r) = 1 - \frac{m(r)}{tc(r)}
\end{align}
where $m(r)$ is the mismatch count and $tc(r)$ is the total number of checks of the generated RTL code $r$. We then select the $K$ candidates with the highest scores, denoted as follows:
\begin{align}
    \mathcal{R}_{i,0}^{\star} =\underset{A \subseteq \mathcal{R}_i,|A|=K}{\arg \max } \sum_{r \in A} s(r)
\end{align}


\noindent3. We employ a debug agent to perform debugging trials, generating a new candidate $r^{\star}_{trial} = D(r^{\star})$ for each selected candidate $r^{\star}$, where $D(r)$ denotes the debug trial for RTL code $r$. The selected candidate set at iteration $m$ is then updated as:

\begin{align}
    R^{\star}_{i,m} = \{ \underset{r\in  \{D(r^\star),r^\star\}} {\arg\max}s(r) \mid  r^\star \in \mathcal{R}^{\star}_{i,m-1} \}
\end{align}
And this process repeats until $\underset{r \in R^{\star}_{i,m} }{max} s(r) = 1$ or the iteration limit is reached.

\subsection{RTL Debugging with State Checkpoint Mechanism}
\label{debugging-details}

Our objective is to design a mechanism that is entirely LLM-based, independent of third-party closed-source tools, and effective in improving RTL debugging efficiency. To achieve this, we propose the RTL Debugging with State Checkpoint Mechanism, which follows these steps:

\noindent 1. Utilizing the optimized testbench generated in Step \circlednumber{1} and \circlednumber{2}, we extract the input vector $\mathbf{I}$, the DUT, and the output $\mathbf{O}$, and subsequently identify the earliest mismatch point as: 
\begin{align} 
t_m = \min \{t \mid \mathbf{O}_{DUT}(t) \neq \mathbf{O}_{exp}(t), t \geq 0 \} 
\end{align} 
where $\mathbf{O}_{\text{DUT}}(t)$ is the output vector from the DUT at time $t$, and $\mathbf{O}_{\exp}(t)$ is the expected output vector at time $t$.

\noindent 2. We generate the State Checkpoint and collect the textual waveform window as: 
\begin{align}
    W = \Big\{ & \big( \mathbf{I}(t'), \mathbf{O}_{DUT}(t'), \mathbf{O}_{exp}(t') \big) \mid \nonumber \\
    & t' \in \big[ \max(t_m - L_W , 0), t_m \big] \Big\}
\end{align}
where $W$ represents the textual waveform window, $L_W$ is the window length parameter, and $\mathbf{I}(t')$ is the input vector at clock edge $t'$.

\noindent 3. The debugging agent then takes the textual waveform window $W$ and the original testbench as inputs, generates a new trial of debugged RTL code, and performs replacement actions to correct the identified faults in the RTL. The simulation is subsequently rerun to verify if the mismatch is resolved in the updated State Checkpoint.



%

\section{Experiments}
\subsection{Experimental Setup}
\noindent\textbf{Benchmarks.} Two widely used benchmarking datasets: VerilogEval-v1-Human\cite{liu2023verilogeval} and VerilogEval-v2\cite{ho2024verilogcoder}. \textbf{Model.} Claude 3.5 Sonnet 2024-10-22 \cite{Claude3}. \textbf{Baselines.} i. Vanilla models: 
Generating RTL code in a single pass using language models, including both general-purpose LLMs (e.g. GPT-4o \cite{gpt-4o}, Claude 3.5 Sonnet\cite{Claude3}) and RTL-specified models (e.g. ITERTL\cite{wu2024itertl}, CodeV\cite{zhao2024codev}). ii. LLM Agent systems:
Agent-based systems designed to enhance LLM performance for RTL code generation, including open-source solutions (OriGen\cite{cui2024origen}) and closed-source ones (e.g. VeriAssist\cite{huang2024towards}, AutoVCoder\cite{gao2024autovcoder}, VerilogCoder\cite{ho2024verilogcoder}, AIVRIL\cite{sami2024aivril}). \textbf{Configurations.}
Our implementation of \nickname integrates the open-source Verilog compiler and simulator Icarus Verilog \cite{iverilog} with an LLM-agnostic API interface offered by the open-source framework LlamaIndex \cite{Liu_LlamaIndex_2022}. Based on the superior performance\cite{swe-bench-sonnet, sami2024aivril, swaroopa2024evaluating} of Claude 3.5 Sonnet \cite{Claude3} in multiple coding tasks, we choose to run experiments with that language model. In accordance with the VerilogEval V1 \cite{liu2023verilogeval} and V2 \cite{pinckney2024revisiting} benchmark, we conducted experiments to measure Pass@1 under 2 settings: \textit{Low Temperature} (T=0, top\_p=0.01, n=1) and \textit{High Temperature} (T=0.85, top\_p=0.95, n=20). Here, temperature T is described in Sec.~\ref{background}. Top\_p limits the output pool to a cumulative probability threshold. n represents the number of evaluation runs. \noindent\textbf{Metrics.}
Following prior works\cite{gao2024autovcoder, huang2024towards, pinckney2024revisiting}, the Pass@1 metric is computed as:
\begin{equation}
    pass@k = \mathbb{E}_{\text{Problems}} \left[ 1 - \frac{\binom{n - c_p}{k}}{\binom{n}{k}} \right]
\end{equation}
where $k=1$, and 
$c_p$ is the number of passing runs. This Pass@1 metric, which accounts for multiple runs, reflects the expected percentage of problems that the system solves correctly when executed once for each problem.





\begin{table}[ht]
\centering
\caption{Pass rates of different temperature configurations in \nickname.}
\label{tab:low_vs_high_temp}
\resizebox{1\linewidth}{!}{
    \begin{NiceTabular}{ccc}[hvlines]
    Config & \Block{1-1}{VerilogEval-Human Pass@1} & \Block{1-1}{VerilogEval-V2 Pass@1} \\
    
    High Temp & \bfseries 94.8 & \bfseries 95.7 \\
    Low Temp  & 89.1 & 93.6 \\
    \end{NiceTabular}
}
\end{table}

As shown in Table~\ref{tab:low_vs_high_temp}, the High Temperature setting achieves higher Pass@1, so it is adopted for subsequent experiments.



\subsection{Key Results}
Table~\ref{tab:pass_rates} shows the comparison of \nickname and baselines.
For a fair comparison, we select the highest pass rate
among their experiment configuration. 
\nickname achieves the best performance on both benchmarks, delivering consistent improvements over specialized (e.g., VerilogCoder) and general-purpose systems(e.g., GPT-4, CodeQwen). Specifically, \nickname obtains a Pass@1 score of 94.8\% on VerilogEval-Human and 95.7\% on VerilogEval-V2, surpassing all baselines. These results underline the effectiveness of \nickname in advancing the capabilities of LLM-driven coding systems.

\begin{table}[t]
\centering
\caption{Pass rates of recent LLMs and Coding Systems. The highest Pass@1 among different temperature settings is reported.}
\label{tab:pass_rates}
\resizebox{1\linewidth}{!}{
    \begin{NiceTabular}{ccccc}[hvlines]
    System & \Block{1-1}{System\\Type} & LLM Model & \Block{1-1}{VerilogEval-\\Human\\Pass@1} & \Block{1-1}{VerilogEval-\\V2\\Pass@1} \\
    
    \Block{3-1}{Generic LLM}
     & N/A & GPT-4o & 51.3 & N/A \\
     & N/A & Claude 3.5 Sonnet$^{*}$ & 60.3 & N/A \\
     & N/A & \Block{1-1}{Claude 3.5 Sonnet\\2024-10-22} & 75.0 & 72.4 \\
    
    \Block{2-1}{RTL-specified\\LLM}
     & N/A & ITERTL\cite{wu2024itertl} & 42.9 & N/A \\
     & N/A & CodeV\cite{zhao2024codev} & 53.2 & N/A \\

    OriGen\cite{cui2024origen} & \bfseries \Block{1-1}{Open\\Source} &  \Block{1-1}{DeepSeek-Coder-7B\\ + LoRA} & 54.4 & N/A \\
    VeriAssist\cite{huang2024towards}
     & \Block{1-1}{Closed\\Source} & GPT-4 & 50.5 & N/A \\
     
     AutoVCoder\cite{gao2024autovcoder}
     & \Block{1-1}{Closed\\Source} & CodeQwen1.5-7B & 48.5 & N/A \\
    
    VerilogCoder\cite{ho2024verilogcoder} 
     & \Block{1-1}{Closed\\Source} & GPT-4 Turbo & N/A & 94.2 \\
     
     AIVRIL\cite{sami2024aivril}
     & \Block{1-1}{Closed\\Source} & Claude 3.5 Sonnet$^{*}$ & 64.7 & N/A \\
    
   \bfseries \Block{2-1}{\nickname (ours)}
     &  \bfseries \Block{2-1}{Open\\Source} & \Block{1-1}{Claude 3.5 Sonnet\\2024-10-22} & \bfseries 94.8 & \bfseries 95.7 \\
      & & Improvement(${\Delta}$)$^{\dagger}$ & +19.8 & +23.3
    \end{NiceTabular}
}
\begin{minipage}{\linewidth}
$^{*}$ Claude 3.5 Sonnet has two versions: 2024-06-20 and 2024-10-22. The cited paper\cite{sami2024aivril} does not specify which version was used.

$^{\dagger}$ The improvement is directly compared to the performance of the same model without employing the \nickname system.
\end{minipage}
\vspace{-8pt}
\end{table}


MAGE not only outperforms closed-source systems but also demonstrates substantial improvement over accessible solutions, including vanilla models and open-source coding systems. For instance, compared to the highest-performing vanilla LLMs (e.g., Claude 3.5 Sonnet), MAGE achieves a relative improvement of +19.8\% on VerilogEval-Human and +23.3\% on VerilogEval-V2. Similarly, against open-source systems like OriGen (54.4\% Pass@1 on VerilogEval-Human), MAGE significantly improves the functional correctness, showcasing its superior performance in scenarios where accessible and reproducible solutions are prioritized.

\subsection{Ablation Study}

\noindent\textbf{Multi-Agent System.} We conducted an ablation study to assess the effectiveness of task distribution among multiple agents. The study compared three configurations: (a) Vanilla LLM, which involves one-pass RTL code generation using a single LLM; (b) Single-Agent, where different agents in the \nickname system were merged into a single agent by sharing a common generation history; and (c) Multi-Agent, the proposed system that assigns tasks to specialized agents based on their roles. As shown in Table~\ref{tab:multi-agent-ablation}, the multi-agent configuration achieves the highest pass rate (93.6\%), outperforming both the vanilla (72.4\%) and single-agent (83.9\%) setups. These results indicate that effective task partitioning can significantly enhance the performance of LLM-based systems, particularly for complex tasks like RTL code generation, which require handling both synthesizable RTL code and non-synthesizable testbenches.

\begin{figure}[t]
    \centering
    \includegraphics[width=\linewidth]{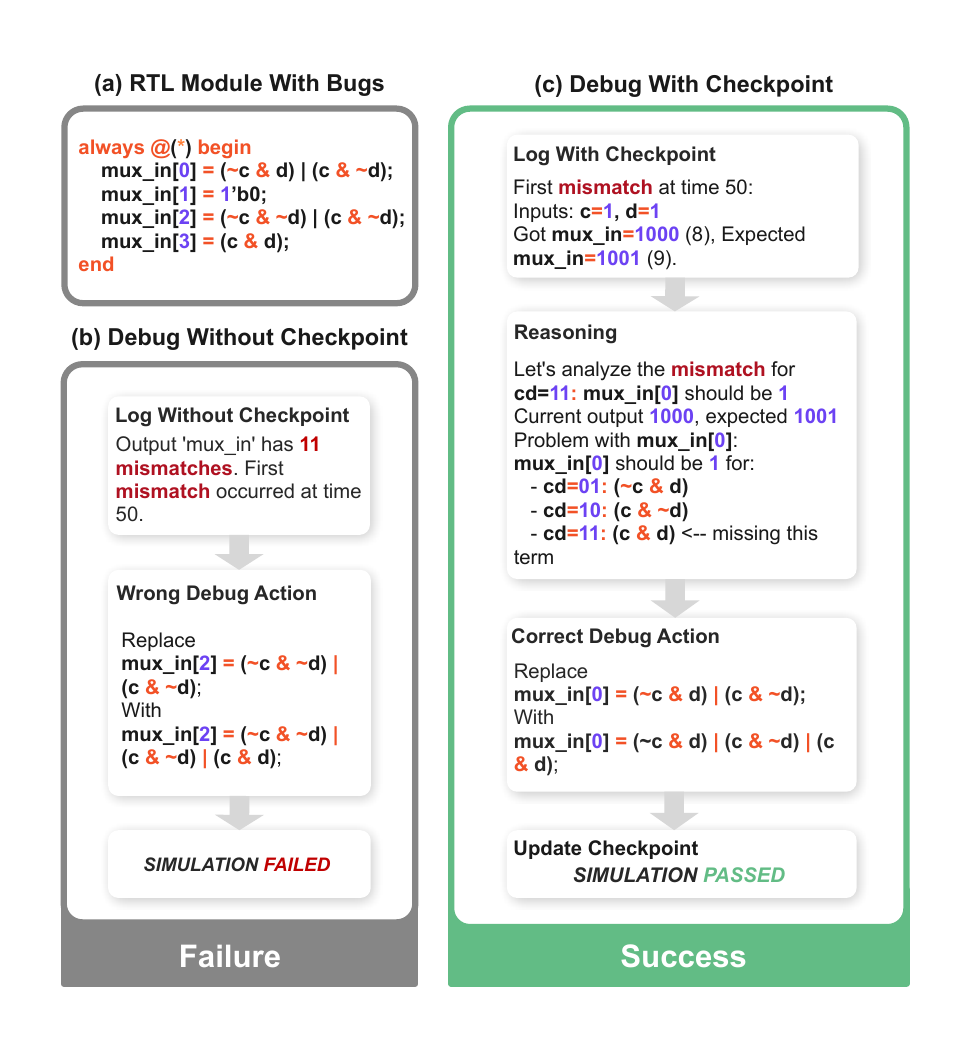}
   
    \caption{The Case Study of RTL Code State Checkpoint on Prob093-ece241-2014-q3.}
    \label{fig:fig-checkpoint-case-study}
    \vspace{-10pt}
\end{figure}

\begin{table}[t]
\centering
\caption{Multi-Agent task distribution ablation study: Pass Rates with Claude 3.5 Sonnet 2024-10-22, Low-Temperature Setting.}
\label{tab:multi-agent-ablation}

\resizebox{0.85\linewidth}{!}{
    \begin{NiceTabular}{ccc}[hvlines]
    Config & Type & \Block{1-1}{VerilogEval-V2 Pass@1} \\
    
    \Block{1-1}{Vanila LLM} & Pass\% & 72.4  \\
    \Block{2-1}{Single-Agent} & Pass\% & 83.9 \\
     & Improvement(${\Delta}$) & \bfseries +11.5  \\
    \Block{2-1}{Multi-Agent} & Pass\% & 93.6 \\
    & Improvement(${\Delta}$) & \bfseries +21.2  \\
    \end{NiceTabular}
}
\vspace{-11pt}
\end{table}

\noindent\textbf{RTL Code State Checkpoint Mechanism.} We also conducted case studies to evaluate the effectiveness of the proposed RTL Code State Checkpoint mechanism. For example, Fig.~\ref{fig:fig-checkpoint-case-study} illustrates a debugging case study with and without state checkpoints. Without checkpoints, when only the log is provided to the LLM Debug Agent, the agent can only approximate the problematic location in the buggy RTL code, making it unlikely to identify and fix the issue. In contrast, with state checkpoints included in the log, the LLM Debug Agent can reason more effectively, pinpointing mismatches in the output and identifying missing logic terms. This leads to a more accurate and efficient bug-fixing process.

\begin{figure}[t]
    \centering
    \includegraphics[width=\linewidth]{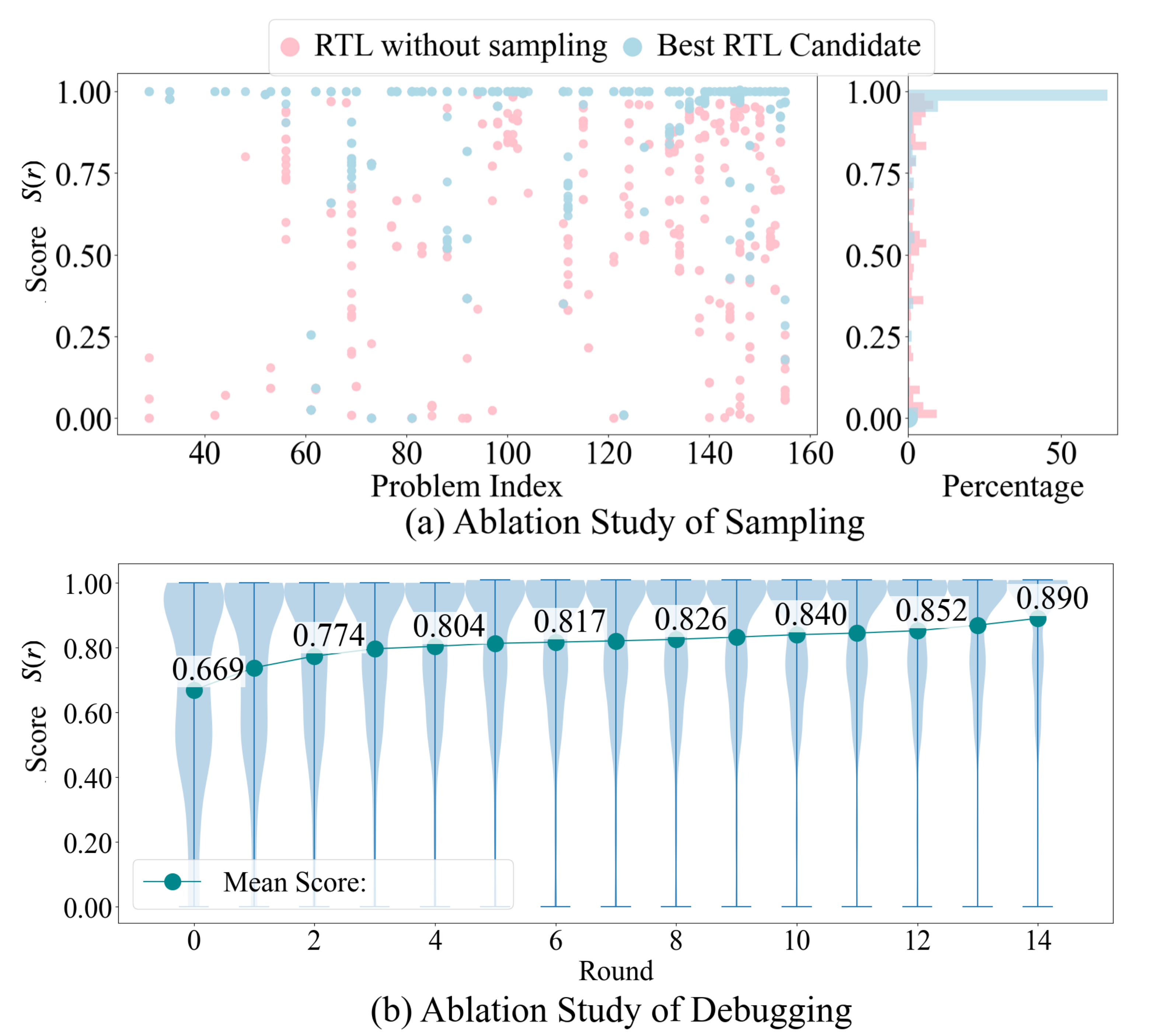}
    \caption{Score $S(r)$ improvement of RTL by sampling and debugging. (a) Score distribution: generated RTL without sampling versus sampled and selected best RTL candidate; (b) Score distribution and the mean score of generated RTL in each debug round. Data of problems fixed before entering the debug stage are not included.}
    \label{fig:fig-ablation-sampling-debugging}
    \vspace{-12pt}
\end{figure}

\noindent\textbf{Sampling and Debugging Mechanisms.} Our experimental data demonstrates the effectiveness of the proposed sampling and debugging mechanisms in enhancing the quality of generated RTL. Fig.~\ref{fig:fig-ablation-sampling-debugging}(a) illustrates that the RTL generated with the sampling strategy consistently outperforms the RTL without sampling across different problems. The score distribution indicates that, without sampling, the RTL scores are nearly uniformly spread across the range $[0, 1]$. However, after applying the sampling method, the scores are concentrated near 1, reflecting a significant quality improvement. Fig.~\ref{fig:fig-ablation-sampling-debugging} (b) depicts the improvement in the mean score across multiple rounds of debugging, starting from an initial score of 0.669 and reaching 0.890 after sufficient refinement. This consistent increase in score demonstrates the cumulative benefit of iterative debugging, leading to more optimal RTL solutions.

\vspace{-10pt}
\section{Conclusion}
In this paper, we propose \nickname, the first open-source LLM-based multi-agent system designed for automated and accurate Verilog RTL code generation. Integrated with a novel High-Temperature RTL Sampling and Scoring process, \nickname 
effectively explores more potentially correct candidates, leading to higher pass rates than prior studies. Augmented with RTL Debugging with State Checkpoint Mechanism, \nickname further optimizes the code with more precise feedback.  Our system represents
a critical step toward automating and optimizing hardware design workflows, offering a more robust methodology for AI-driven RTL design.

\bibliographystyle{IEEEtran}
\bibliography{IEEE_conference}

\end{document}